\documentclass[]{pasj01}
\usepackage{lineno}
\draft 
\Received{$\langle$reception date$\rangle$}
\Accepted{$\langle$acception date$\rangle$}
\Published{$\langle$publication date$\rangle$}

\begin{document}

\title{Gaia DR2 and EDR3 data and evolutionary  status of post-AGB stars with high radial velocities}

\author{Wako \textsc{Aoki}\altaffilmark{1} \altaffilmark{2}}
\altaffiltext{1}{National Astronomical Observatory, 
 2-21-1 Osawa, Mitaka, Tokyo 181-8588, Japan }
 \altaffiltext{2}{Department of Astronomical Science, School of Physical Sciences, The Graduate University of Advanced Studies (SOKENDAI), 2-21-1 Osawa, Mitaka,
Tokyo 181-8588, Japan}
\email{aoki.wako@nao.ac.jp}

\author{Tadafumi \textsc{Matsuno}\altaffilmark{3}}
\altaffiltext{3}{Kapteyn Astronomical Institute, University of Groningen  \\ Landleven 12, 9747 AD Groningen, The Netherlands}
\email{matsuno@astro.rug.nl}

\author{Mudumba \textsc{Parthasarathy}\altaffilmark{1} \altaffilmark{4} \altaffilmark{5}}
\altaffiltext{4}{Indian Institute of Astrophysics, II Block, Koramangala, Bangalore 560 034, INDIA}
\altaffiltext{5}{Department of Physics and Astronomy, Vanderbilt University, Nashville, TN 37235, USA}

\email{m-partha@hotmail.com}



\maketitle

\begin{abstract} 
 Using the {\it Gaia} DR2 and EDR3 data and list of post-AGB
 candidates, we  investigate the parallax, proper motion and
   binarity for twenty post-AGB stars and candidates having high
   radial velocities.  From their {\it Gaia} distances their
 luminosities and kinematics are derived.  The evolutionary status of
 these stars is discussed from their location on the post-AGB
 evolutionary tracks.  Nine stars are confirmed to be post-AGB stars
 that have their initial main-sequence mass around one or two solar
 masses. From their kinematics information, two objects among them are
 identified to clearly belong to the halo population, suggesting that
 low-mass. We discuss on the origin and evolutionary status of other
 objects in the sample of this work with high radial velocities.
\end{abstract}

\section{Introduction}

Post-AGB stars are transition objects evolving from the tip of the AGB
horizontally towards the left in the H-R diagram into early stages of
young planetary nebulae (PNe).  The post-AGB evolutionary stage is
short-lived, depending on the core-mass \citep{schoenberner83,iben83}. During
the transition from the tip of the AGB to early stages of young PNe
phase they appear as M-,K-, G-, F-, A-, and OB-type post-AGB
supergiants for a short period \citep{partha86, partha89, pottasch88, partha93a,partha93b}. They
mimic the spectra of supergiants because of their extended thin
atmospheres around the white-dwarf like C-O core (after severe
mass-loss and the termination of the AGB phase of evolution).  Before
the advent of IRAS satellite, very few post-AGB supergiant candidates
were known. Analysis of IRAS data has revealed many cool to hot
post-AGB supergiants \citep{preite-marthinez88,kwok89}.
The list of post-AGB stars detected from the analysis of IRAS data by
several investigators can found in the paper of \citet{vickers15} and  references therein.
Progresses in understanding of post-AGB stars are found in review papers 
(e.g., \cite{vanwinckel03, kamath22u}).

Multi-wavelength studies of significant sample of post-AGB candidates
were carried out by several investigators during the past 35 years
which enabled us to understand their chemical composition, circumstellar
shells, s-process nucleosynthesis and late stages of evolution of low-mass 
stars (e.g., \cite{desmedt12,desmedt16,kamath22,partha22}, and references therein). However, for a better understanding of these stars,  their
distances, radial velocities and accurate proper-motion measurements
are required.  With the advent of {\it Gaia} (DR2 and EDR3; \citet{gaia18,lindegren18}, accurate  parallaxes(distances), radial velocities and proper-motions measurements of a
large sample post-AGB stars became available.

 In an earlier paper
\citep{partha20}, we  studied the Gaia DR2 data and
evolutionary status of eight high velocity hot post-AGB stars. Finding high-velocity objects among post-AGB stars is useful to constrain the final stage of stellar evolution of low-metallicity, low-mass stars in the halo population. 
Such stars are very  rare and are old low-mass stars in advanced stages of  evolution.  Some
of them may belong to the Galactic halo.
In this paper we present an analysis of {\it Gaia} DR2 and EDR3 data of
stars listed as post-AGB stars or candidates in literature.


\section{Data and analysis}

We investigate the list of stars given in the paper of \citet{vickers15}  as likely or possibly post-AGB stars, and searched the {\it Gaia} DR2
and EDR3 for post-AGB stars with radial velocities and with accurate parallaxes. The sample of \citet{vickers15} contains the list of almost all the known post-AGB stars. Here we define
post-AGB stars as those which are in the transition region between the tip of the AGB and very early stages of
planetary nebulae (PNe).  These objects are often termed as proto-planetary nebulae.

We select 20 objects that have absolute values of radial velocities larger than 45~km~s$^{-1}$ ($|RV|>45$~km~s$^{-1}$).
The galactic longitudes and latitudes, parallaxes, radial velocities,
$G$ ({\it Gaia G} band magnitude), $V$, $(B-V)$ and Spectral types are given in
tables \ref{tab:obj} and \ref{tab:param}.  Spectral types, $V, B$ magnitudes are taken from SIMBAD.
Among the twenty stars twelve are high galactic latitude stars, eleven
stars have high negative radial velocities and nine have high positive
radial velocities.

The parallaxes taken from  {\it Gaia} DR2 and EDR3 are given in table~\ref{tab:param}.
  The distance derived from the parallaxes are also listed in the table. 
  Four stars have a large relative parallax uncertainty ($>$20\%) in {\it Gaia} EDR3. For these stars, only the lower limit of the distance is presented. The table also gives the distances estimated by \citet{Bailer-Jones21} using a prior constructed from a three-dimensional model of the Galaxy. Excluding the above four objects, the two estimates of the distance for each object agree within 5\%. We adopt the distances simply obtained from the parallaxes in the present work.
  
The normalized unit weight error (RUWE) values are also given in the
  table. The RUEW values also indicate reliability of the parallaxes. Whereas RUWE values are sensitive to the photocentric motions of unresolved companions \citep{lindegren21, stassun21}, they could be affected by other factors including nebulosity of proto planetary nebulae. 
The RUWE values of nine stars, including the above four
objects with large uncertainties of parallaxes, are larger than 1.4.
Among them, four objects are likely to be
  post-AGB stars according to previous studies on stellar properties
  including infrared excess, and metal depletion. See below more details
  on these stars. The remaining five stars have large uncertainties in
  parallaxes and/or very large RUWE values, and, hence, are not
  regarded as candidates of post-AGB stars in this paper.
   
The other eleven stars in our sample have RUWE values smaller than 1.4. Among them, BD+33 2642 is suggested to belong a binary system
  \citep{vanwinckel14}.
For the remaining ten objects, there is no signature of binarity from the Gaia astrometry.

The kinematic information that is calculated based on the {\it Gaia} EDR3 astrometry is presented in table 3, excluding the four objects with large uncertainties of parallaxes.


$E(B-V)$ values are obtained from dust maps in literature or by comparing expected intrinsic colors with observed ones (table \ref{tab:obj}).
We use three-dimensional dust extinction map from \citet{chen19} and \citet{green18} and two-dimensional dust extinction map from \citet{schlegel98}.
For a star to have an $E(B-V)$ estimate from three-dimensional maps, it needs to have a precise parallax measurement (relative uncertainty smaller than 20\%) and be in the sky coverage of the maps.
Since \citet{chen19} focus on low Galactic latitude field ($|b|<10^\circ$), we prioritize values from \citet{chen19} over \citet{green18} for objects with $|b|<10^\circ$. 
We note that \citet{green18} only covers the sky with declination larger than $-30^\circ$ and hence we could not derive $E(B-V)$ from three-dimensional maps for HD 16745 and HD 178443 despite precise parallax measurements available for these objects.
The extinction coefficients from \citet{green18} and \citet{chen19} are converted to $E(B-V)$ using values provided in \citet{green18},
\citet{schlafly11}, and \citet{casagrande19}.  
In addition to the interstellar extinctions considered in these dust maps, some objects could be affected by circumstellar dust extinction given the evolutionary status of the objects. 
Thirteen objects are indeed IRAS sources
and their $(B-V)$ colours are likely affected by circumstellar reddening. 
For instance, the $E(B-V)$ of HD~56126 (IRAS~07134+1005) estimated from the spectral type is 0.56, whereas the the $E(B-V)$ from the dust map is quite small (0.08 or less).
  For these stars we used the
observed $(B-V)$ values from SIMBAD and intrinsic $(B-V)_{0}$ values
estimated from their spectral types using Table 15.7 of Allen's Astrophysical Quantities \citep{cox00} with interpolation to derive $E(B-V)$ values. 
The $E(B-V)$ values derived in this way are prioritized over the values from dust maps.

On the other hand, the $E(B-V)$ values of IRAS~07140-2321, IRAS~07227-1320, and IRAS~14325-2321 estimated from the spectral types are significantly smaller than those from the dust map.  This suggests that the estimate of the reddening from the dust map or spectral types could be uncertain for these objects. For these three stars, we adopt $E(B-V)$ from the dust map.

The typical errors of $E(B-V)$ from the dust map is 0.02--0.03 for high Galactic latitude objects. We adopt 0.05 as the uncertainty of reddening of these objects to determine the luminosity. This is consistent with the error of $E(B-V)$ given in \citet{vickers15} for our sample (0.043 on average). For objects with large reddening, in particular objects that could be affected by circumstellar reddening, we assume the error of $E(B-V)$ to be 0.2, including the uncertainty of the subclass of spectral types that results in difference of $E(B-V)$. The $E(B-V)$ and the error adopted are given in table 1.  Taking the errors into account, the $E(B-V)$ values estimated in this study agree well with those obtained for HD~56126 (0.43) and IRAS~14325--6428 (1.07) by \citet{kamath22}. Although our values are slightly larger than those for IRAS~05208--2035 (0.01) and HD~46703 (0.23) by \citet{oomen18}, and for IRAS~08187--1905 (0.07) and HD~161796 (0.13) by \citet{kamath22}, the discrepancy is smaller than 0.1. if the error ranges are taken into account.

The absolute $V$ magnitudes are calculated from the apparent magnitudes and distances given in tables 1 and 2, respectively. The luminosity is calculated from the absolute magnitude and the bolometric
corrections taken from \citet{cox00}. The values
of bolometric corrections in \citet{flower96} are 0.1--0.25 mag larger
than those of \citet{cox00}, resulting in differences in $\log (L/L_{\odot})$ of
less than 0.1 dex. The changes of $E(B-V)$ of 0.05 and 0.2 result in the difference of $\log(L/L_{\odot}$) of 0.08 and 0.25, respectively.  The $T_{\rm eff}$ values and spectral
types of most stars are available from the literature. These values are given in Table~2. 

\begin{figure}
\begin{center}
  \includegraphics[width=\textwidth]{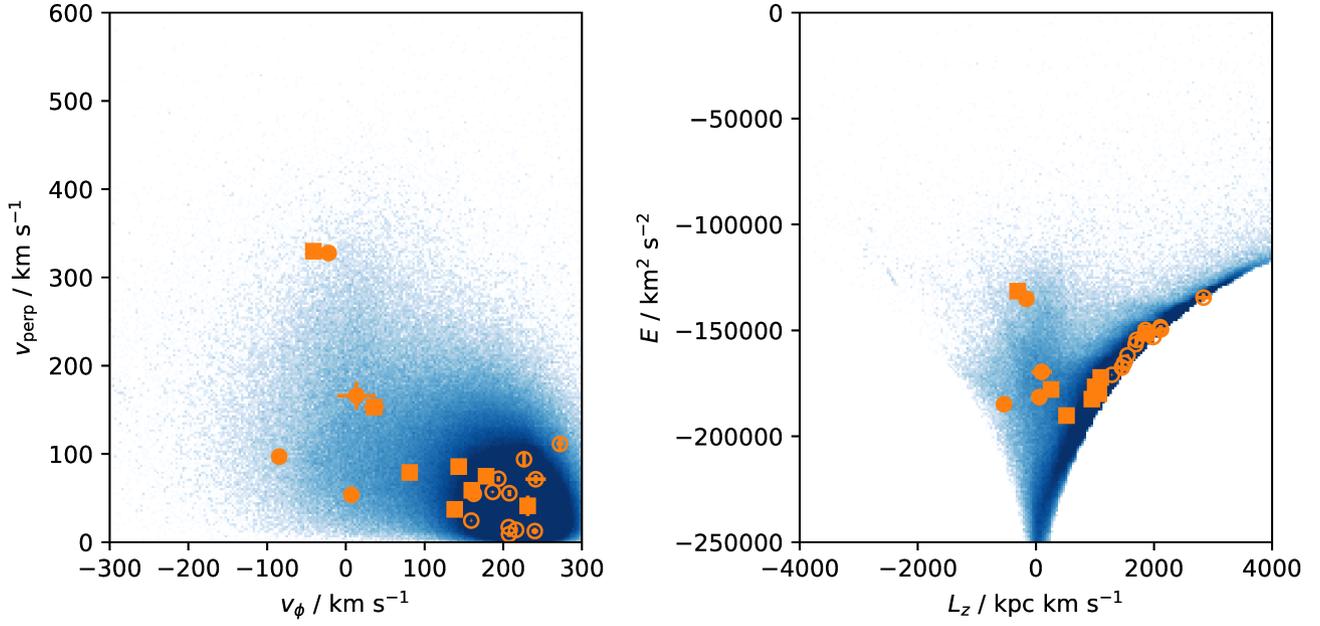}
\end{center}
\caption{Kinematics of the post-AGB candidates in the present study in the Galactocentric frame.  The four objects with halo kinematics found by the present work are shown by filled circles, whereas those with disk-like kinematics are shown by open circles. The filled squares are the objects included in \citet{partha20}.}\label{fig:kinematics}
\end{figure}

Kinematics are calculated using the {\it Gaia} EDR3 parallaxes and
proper motion measurements. We adopt 8.21 kpc as the distance between
the Sun and the Galactic center \citep{mcmillan17} and 0.021 kpc as the
vertical offset of the Sun \citep{bennett19}. Solar motion is
adopted from \citet{schonrich10} for radial and vertical
velocities (11.1~km~s$^{-1}$ and 7.25~km~s$^{-1}$, respectively) and is
calculated as 245.34~km~s$^{-1}$ using proper motion measurements by Reid
and \citet{brunthaler04}. 
The orbital energy is calculated assuming the Milky Way potential from \citet{mcmillan17}.
These results are given in table 3. Kinematics of the post-AGB star candidates in Galactocentric frame are presented in figure\ref{fig:kinematics}.

We note that the sample selection of high velocity post-AGB stars
would not be affected if they belong to low mass binaries because the
radial velocity variations expected for low mass binaries are not as
large as the radial velocities of the stars studied in this paper
(tables 1 and 2).


\section{Notes on the  twenty high velocity post-AGB  candidates}

\begin{figure}
 \begin{center}
  \includegraphics[width=10cm]{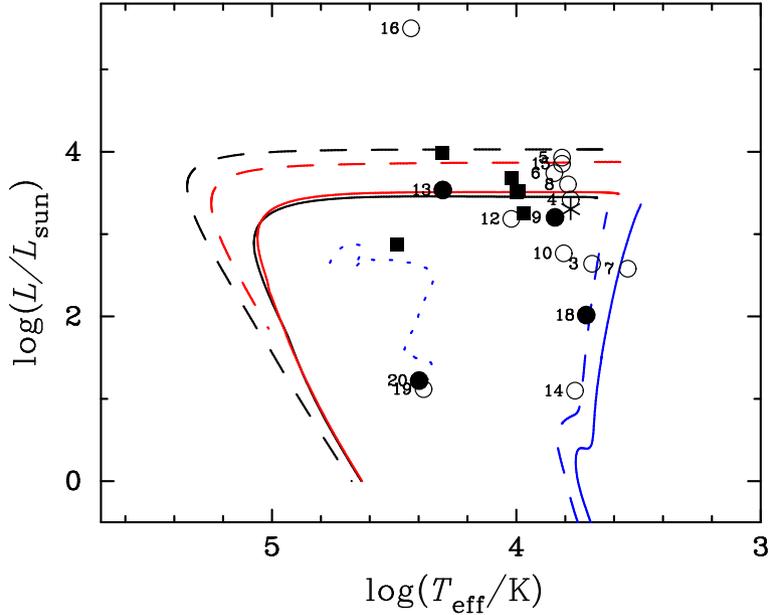}
 \end{center}
 \caption{
   Evolutionary tracks of post-AGB phases taken from
   \citet{miller16} for initial masses of 1.0~M$_{\odot}$ (solid
   lines) and 2.0~M$_{\odot}$ (dotted lines) with $Z=0.02$ (red) and
   $Z=0.001$ (black). The evolutionary track of post-HB
   star for the core mass of 0.52~M$_{\odot}$ with [Fe/H]$=-1.48$
   taken from \citet{dorman93} is shown by dotted (blue) line. The isochrones of Yonsei-Yale models for the age of 9 Gyr are shown by (blue) solid and dashed lines for [Fe/H]=0.0 and $-1.7$, respectively, presenting the red giant branches. The four objects with halo kinematics found by the present work are shown by filled circles, among which HD~116745 and BD+33$^{\circ}$2642 are clearly post-AGB stars.  Other 12
   objects with reliable luminosity are shown by open
   circles. The object numbers given in the tables are presented. The five objects reported in Figure 1 of \citet{partha20} are plotted by filled squares. The post-AGB star BD$+14^{\circ}3061$ is plotted by asterisk, adopting $T_{\rm eff}=6000$~K \citep{bond20}. }\label{fig:track}
\end{figure}

Figure~\ref{fig:track} shows the luminosity of the objects as a function of effective temperature with the post-AGB evolution
tracks \citep{miller16}. The four objects with large uncertainties in the parallaxes (see \S~2) are excluded. This figure indicates that typical luminosity range of low-mass post-AGB stars is $3<\log(L/L_{\odot})<4$.

We find that nine objects have luminosity of this range, among which five have reliable Gaia parallaxes (the RUWE values are smaller than 1.4). BD$-12$ 4970 that has very high luminosity ($\log(L/L_{\odot})=5.5$) is also regarded as a post AGB star. Among the remaining ten stars with higher or lower luminosity than the above post-AGB star range, five with small RUWE have lower luminosity $\log(L/L_{\odot})<3$ and the other five have large RUWE values.


Here we report some detailed information for individual objects separately with the above grouping. 
It should be noted that the luminosities of binary stars are still uncertain due to the uncertainty of parallaxes. Although most of them are found in the groups with large RUWE values in this section, some known binary stars are also included in the groups with small RUWE values. Information on the binarity is given in the following notes for individual objects when available.

\subsection{Post-AGB stars with small RUWE values}

\begin{itemize}

\item{HD 56126 (IRAS 07134+1005)}

  It is a high Galactic latitude,  and high velocity, metal-poor, F-type post-AGB star with 21-micron emission feature.  It is overabundant in carbon and s-process elements \citep{partha92}. \citet{desmedt16} report detailed abundances including [Fe/H]$=-0.91$ and large excesses of s-process elements (e.g., [Ba/Fe]$=1.82$). More recently, \citet{kamath22} list this object as a single post-AGB star with s-process enrichment.

\item{ IRAS 07140-2321 (V421 CMa)}

 \citet{gielen11} derived $T_{\rm eff}$ = 7000~K, $\log g$ = 1.5, and [Fe/H] = $-0.8$. 

\item{ HD 116745 (CD$-46^{\circ} 8644$, Fehrenbach’s star, ROA 24,)}

  It is a high galactic latitude and high velocity metal-poor halo post-AGB star \citep{gonzalez92}. They found  it to be  overabundant in carbon and s-process elements. They derived $T_{\rm eff}$ = 6950~K, $\log g$ = 1.15 and [Fe/H] = $-1.77$. It is a member of the globular cluster Omega Cen.

\item{ IRAS 17436+5003 (HD 161796)}
  
 It is a high galactic latitude and high velocity   F-type post-AGB supergiant  \citep{partha86}. \citet{luck90}  derived $T_{\rm eff}$  = 6500~K, $\log g$ = 0.70, and [Fe/H] = $-0.32$. This object is listed as a single post-AGB star without s-process enrichment by \citet{kamath22}.

\item{ BD$-12^{\circ} 4970$ (LS IV -12 13)}

It is a high velocity hot (B0.5Ia) post-AGB candidate. It is not an IRAS source. High resolution  spectroscopic study of this star is important.

\item{ BD+33 2642}

  It is a high Galactic latitude and high velocity, metal-poor  hot post-AGB star. This object is also classified into a proto planetary nebula. However, its nebula is not bright, and it is not an IRAS source.  \citet{napiwotzki94} studied this star and derived $T_{\rm eff}$  = 20,000~K, $\log g$ =2.9,  and [Fe/H]  = $-2.0$. Its chemical composition indicates depletion of  refractory elements. The [O/H]= $-0.8$ indicates it is intrinsically metal-poor. This star belongs to a binary system with a low-mass faint companion. High resolution spectrum shows no spectral  features of the secondary. The orbital period determined by \citet{vanwinckel14} is 1105 days. The binarity of this object would not affect the RUWE value, which is smaller than 1.4 (1.295).

\end{itemize}

\subsection{Post-AGB stars and candidates with large RUWE values}

\begin{itemize}

\item{HD 46703 (IRAS 06338+5333)}

This star is a high Galactic latitude, and  high velocity, metal-poor F-type pop II post-AGB star \citep{luck84}. \citet{luck84} derived  $T_{\rm eff}$ = 6000~K, $\log g$ = 0.4, [Fe/H] = $-1.57$. \citet{partha86} were the first to find that it is a weak IRAS source with far-IR colours and flux distribution similar to that of high Galactic latitude post-AGB star HD 161796 \citep{partha86}. \citet{hrivnak08} also report [M/H]$=-0.6$ for this object, discussing depletion and binarity. This belongs to a binary system. \citet{oomen18} report the orbital period of 597 days for this object. The RUWE value of this star is 1.622, which is clearly higher than 1.4 as expected from the binarity.
  
\item{IRAS 08187-1905 (HD 70379, V552 Pup)}

 It is a high galactic latitude and high velocity F6 post-AGB supergiant \citep{reddy96}. \citet{reddy96} derived chemical composition of this star from an analysis of high resolution spectra. They derived $T_{\rm eff}$ = 6500~K, $\log g$ = 1.0, [Fe/H] = $-0.5$. This star is listed as a single post-AGB star without s-process enrichment by \citet{kamath22}. The RUWE value of this object is 1.695, which is even higher than that of HD~46703. Although this would not indicate that this object belongs to a binary system, further investigation on the binarity will be useful.

\item{ IRAS 14325-6428}

It is a high velocity F5I star with IRAS colours and flux distribution similar to post-AGB stars and PNe.  \citet{desmedt16} report [Fe/H]$=-0.56$ with excesses of s-process elements. This star is also regarded as a single post-AGB star without s-process enrichment by \citet{kamath22}. The RUWE value of this object is quite large (2.181). Whereas further study to investigate the binarity would be useful, this star can be treated as a post-AGB star with excesses of s-process elements according to literature.

\item{ HD 137569 (IRASF 15240+1452)}

 It is a  high Galactic latitude and high velocity post-AGB star. No Fe lines are detected in its spectrum by \citet{martin04} and \citet{martin06}, who classified it as a metal-poor, hot post-Horizontal Branch (post-HB) star.

\end{itemize}

\subsection{Objects with low luminosity with small RUWE}

\begin{itemize}

\item{ IRAS 07227-1320}

 This star is listed as a possible post-AGB star by \citet{vickers15}. Its spectral type is M1I. No chemical composition study is available. The luminosity of this object ($\log(L_{*}/L_{\odot})=2.58$) is lower than post-AGB stars in our sample.

\item{ BD+32 2754}

This star is also listed as a possible post-AGB star by \citet{vickers15}.  It is a high Galactic latitude and high velocity F-type star. There is no chemical composition analysis of this star. It may belong to Galactic halo. The luminosity of this object ($\log(L_{*}/L_{\odot})=1.10$) is clearly lower than post-AGB stars.

\item{ HD 178443 (LSE 182)}

   It is not an IRAS source. It is a high Galactic latitude and high velocity (343.5 km~s$^{-1}$) star. \citet{mcwilliam95} derived  $T_{\rm eff}$  = 5180~K,  $\log g$ = 1.65, [Fe/H] = $-2.07$.  They classify it as a red-HB star. It is a Galactic halo star (see next section). The luminosity of this object ($\log(L_{*}/L_{\odot})=1.99$) is lower than post-AGB stars in our sample.

\item{ PHL 1580}

     It is a high Galactic latitude and high velocity hot post-AGB star.  \citet{mccausland92} derived  $T_{\rm eff}$  = 24,000~K, $\log g$ = 3.6, [Fe/H] =$-0.6$.  They find it to be carbon deficient. This star may have left the AGB before the third dredge up. The luminosity of this object ($\log(L_{*}/L_{\odot})=1.12$) is clearly lower than post-AGB stars.

\item{LS III +52 5}

  It is a high velocity ($-232.8$ km~s$^{-1}$)  and high proper motion star. In the LS catalogue  its spectral type is given as  OB- \citep{hardorp64}.  It is not an IRAS source. Detailed spectroscopic  study of this star is important. The luminosity of this object ($\log(L_{*}/L_{\odot})=1.22$) is clearly lower than post-AGB stars.

\end{itemize}
  
\subsection{Others with uncertain luminosity and large RUWE}

\begin{itemize}

\item{IRAS 02143+5852}

It is a high radial velocity  F7Iae star. The H$\alpha$ line is in emission.  $T_{\rm eff}$ is estimated from its spectral type to be 6000K. \citet{fujii02} made $BVRIJHK$ photometry.  \citet{omont93} classified it as a carbon-rich post-AGB star.  The error in parallax is large (tables 1 and 2).

\item{IRAS 05089+0459}

  It is a high galactic latitude and high velocity M3I post-AGB candidate. \citet{iyengar97} made near-IR photometric observations ($R$ = 12.68, $I$ = 11.62). There is no chemical composition analysis of this star. The error in {\it Gaia} EDR3 parallax is high (tables 2).

\item{ IRAS 05208-2035 (BD$-20^{\circ} 1073$, AY Lep)}

It is a high Galactic latitude and high velocity post-AGB candidate. \citet{gielen11} derived $T_{\rm eff}$  = 4000~K, $\log g$ = 0.5, and [Fe/H] =0.0. The observed $B-V$  colour indicates that it may be a G-type star. The spectral type is not available in SIMBAD. On the other hand, [Fe/H]$=-0.7$ and small overabundance of s-process elements are derived by \citet{rao12}. \citet{oomen18} derive the orbital period of this binary system to be 23 days. Although the luminosity of this star is still uncertain, it is likely to be a binary post-AGB star.

\item{ IRAS 15210-6554}

  From {\it Gaia} DR2 data we find it  to be a high velocity star. Its spectral type is K2I and Galactic latitude $b$ is $-7.7$ degrees. Based on the IRAS colours and flux distribution it is classified as a post-AGB  star. This star does not have accurate {\it Gaia} DR2 parallax.

\item{ IRAS 18075-0924}

    It is a high velocity star.  Gaia DR2 parallax is not accurate.  Spectroscopic and photometric  study of this star is needed. Based on IRAS colours and flux  distribution it is classified  as a post-AGB  candidate.

\end{itemize}


\section{Discussion and concluding remarks}\label{sec:discussion}

\subsection{Populations of post-AGB stars with high radial velocities}

Nine objects in our sample, HD 46703, HD 56126, IRAS 07140--2321, IRAS
08187--1905, HD 116745, IRAS 14325--6428, HD 137569,
BD$+33^{\circ} 2642$, and IRAS 17436+5003 are identified as post-AGB
stars with high radial velocities (tables 1 and 2).  The very
  luminous object BD-12 4970 is discussed separately.
Their computed
absolute luminosities and comparisons with post-AGB evolutionary track
(Figure 2) indicates that their initial main-sequence mass is less
than 2 solar masses. Among them, only two stars, HD 116745 and
BD$+33^{\circ} 2642$, clearly belong to the galactic halo population
(table 3, figure 1). IRAS 07140-2321 has the largest $L_{z}$ and $E$
(table 3). The other six post-AGB stars are not separated from disk
stars in figures 1 and 2, although they have relatively high radial
velocity.  This indicates that the criterion of the radial velocity
($|V_{\rm Helio}|>45$~km~s$^{-1}$) is not sufficient to effectively
select halo post-AGB stars. The radial velocities of the clear
examples of halo objects identified by this work, HD~116745 and
BD$+33^{\circ} 2642$, are $V_{\rm Helio}=240$ and $-94$ km~s$^{-1}$,
respectively. It should be noted that the above criterion is adopted
in this work as we do not miss halo objects from the sample of
\citet{vickers15}.

IRAS~07140-2321 is a unique object that has high total energy of the orbital motion and high $z$-component of angular
momentum. The star seems to belong to the disk population rather than the halo from the prograde rotation with small $v_{R}$ and $v_{z}$. The distance and the high total energy suggests that it is a outer disk object. 

BD$-12^{\circ} 4970$ (LS IV -12 13) is a hot high velocity star
with accurate parallax. Its computed absolute luminosity indicates
that its initial main-sequence mass may be 4.0 solar masses. 
The kinematics of this object suggest this object belongs to disk population.

Among the objects studied in \citet{partha20}, three objects (LS~3593, LSE~148, and HD~214539) have clear kinematics features of halo objects (figure 1), whereas those of three other stars are not distinguished from disk stars. Another object, LS~5107, has high total energy of the orbital motion and high $z$-component of angular
momentum, as found for IRAS~07140-2321 in the current sample. As LSE~148 is less luminous objects, the clear halo post-AGB stars identified by the study is LS~3593 and HD~214539.

\subsection{Comments for other objects}

The two less luminous stars HD 178443 and PHL 1580 also belong to the halo
population (figure 1).
The high velocity hot metal-poor star
PHL 1580 with accurate parallax is found to have very low 
luminosity (table 2) compared to post-AGB stars.  It may be a hot sub-dwarf star. LSE 182
(HD 178443) is a high velocity metal-poor star in the Galactic halo
and could be a red HB star (McWilliam et al. 1995).

BD$+32^{\circ} 2754$  also has low absolute luminosity. It may
be a sub-dwarf.  IRAS 07227--1320 with the spectral type M1 may be a cool
post-AGB star. It needs further study to understand its chemical
composition and evolutionary stage.  The computed absolute luminosity
of post-AGB star HD 161796 (tables 2) and its location in figure
2 indicates that its initial main-sequence mass may be in the rage around 2
solar masses.  

\citet{kamath15} found dusty post-red giant branch (post-RGB ) stars in LMC and SMC. They found that these stars have mid-IR excesses and stellar parameters  ($T_{\rm eff}$, $\log g$, [Fe/H]) 
similar to those of post-AGB stars, but their luminosities are less than 2500 L$_{\odot}$. Their lower luminosities indicate they have lower masses and radii. Some of the stars
in our sample also have luminosities less than 2500~L$_{\odot}$ (table 2, figure 2) and they may be
post-RGB stars similar to those found by \citet{kamath15}. The very low luminosity
stars like PHL 1580 mentioned above is a puzzle. They may be post-HB stars or evolving
towards AGB-manque star stage.

Recently, \citet{bond20} found BD$+14^{\circ} 3061$ to be a luminous, metal-poor, yellow post-AGB supergiant star in the galactic halo.
He found it to be a a very high-velocity star moving in a retrograde Galactic orbit.  It is not an IRAS source.
The Galactic halo post-AGB stars have relatively low core-mass. They evolve slowly and, by the time they
evolve to G and F-type post-AGB stage, their circumstellar dust shells get dispersed into the interstellar
medium. They never become PNe. The galactic halo post-AGB supergiants are very rare. Discovering them is a challenging task.
\citet{bond20}  derived absolute visual magnitude $M_{V}$ of this star from {\it Gaia} DR2 parallax to be $M_{V} = -3.44$.
Since its bolometric correction is close to zero (i.e., $M_{V} = M_{\rm bol}$), \citet{bond20}  proposed that these Galactic halo
A and F-supergiants  are useful as standard candles as they are luminous and have same
absolute luminosity. Some of the galactic halo post-AGB stars in our sample seems to be similar to BD$+14^{\circ} 3061$.
Extensive survey is needed to detect more galactic halo post-AGB supergiants.

\section{Summary}

This paper investigates the list of post-AGB star candidates of \citet{vickers15} selecting objects with high radial velocities. We identify two clear examples of high-velocity low-mass post-AGB stars and a few candidates from the evolutionary status and kinematics information derived from the {\it Gaia} DR2 and EDR3. Through the studies of this paper and of the previous one \citep{partha20},  four clear halo post-AGB stars are identified (HD~116745, BD+33$^{\circ}$2642, LS~3525 and HD~214539). 
We also find that the list of \citet{vickers15} include objects which are not classified into post-AGB stars, taking the new estimate of luminosity based on parallax measurements with {\it Gaia}. Further studies of the sample of \citet{vickers15} with spectroscopy to determine radial velocities are useful to obtain statistics of post-AGB stars as well as information on individual objects.

\begin{ack}

MP was supported by the NAOJ Visiting Fellow Program of the Research
Coordination Committee, National Astronomical Observatory of Japan
(NAOJ), National Institutes of Natural Sciences(NINS).
\end{ack}


\clearpage

\scriptsize
\begin{table*}
  \tabcolsep 4pt
\tbl{Basic data of  twenty high velocity  post-AGB  candidates	
\label{tab:obj}}{
\begin{tabular}{lrrlrrrrrrrrrrl}
\hline\noalign{\vskip 3pt}
star &  $l$ & $b$ &  Sp. &  $V$ &  $G$ & $ (B-V)$ & $(B-V)_{0}$  & \multicolumn{4}{c}{$E(B-V)$} & $T_{\rm eff}$ &  B.C & Ref. \\
\cline{9-12}
     &  (deg) & (deg) &  &  &  &   & Sp & Sp & SFD &  3D & adopted & (K) &  & \\
\hline\noalign{\vskip3pt}
1)IRAS 02143+5852  &          133.8  & -1.93  &  F7Ie  & 13.8   & 13.51  & 1.22  &  0.48  & 0.74     &  1.05       &         &  ...   & 6000  & -0.07 & 1,2,3\\      
2)IRAS 05089+0459  &          196.3  & -19.5  &  M3I   & 14.08  & 13.13  & 1.74  &        &          &  0.14       &         & ...     &3200  & -2.24 &1,4  \\    
3)IRAS 05208-2035  &          222.8  & -28.3  &        &  9.48  & 8.98   & 1.04  &  0.7:   & 0.3:     &  0.06      &  0.08   & 0.3$\pm0.2$ &4900  &  -0.33 & 5,6,7 \\             
4)HD 46703 &   162.0  & +19.   &  F7I   &9.04    & 8.84   & 0.48  &  0.02  & 0.46     &  0.08       & 0.10    & 0.46$\pm0.20$ & 6000  &   -0.06 & 1,7,8,9\\     
(IRAS 06338+5333)  &&&&&&&&&&&& & \\
5)HD 56126 &  206.7  & +10.0  &  F5Ia  & 8.32   & 8.06   & 0.88  &  0.32  & 0.56     &  0.08       & 0.00    & 0.56$\pm0.20$ &  6500  &   -0.03 & 1,10,11,12 \\      
(IRAS 07134+1005)  &&&&&&&&&&&& & \\
6)IRAS 07140-2321           & 236.6  & -5.4   &  F5I   & 10.73  & 10.49  & 0.43  &  0.23  & 0.20     &  0.59       & 0.57    & 0.57$\pm0.20$ &  7000  &    0.0 & 1,13 \\
7)IRAS 07227-1320           & 228.7  & +1.2   &  M1I   & 12.55  & 11.6   & 1.96  &  1.69  & 0.27     &  0.51       & 0.43    & 0.43$\pm0.20$ &  3500  &  -1.45 & 1,14 \\
8)IRAS 08187-1905           & 240.6  & +9.8   &  F6Ib/II&  9.02 & 8.83   & 0.61  &  0.40  & 0.21     &  0.11       & 0.15  & 0.21$\pm0.05$ &  6150  &  - 0.06 & 1,12,15 \\
9)HD 116745                 & 309.1  & +15.2  &  A7/A9e & 10.79 & 10.68  & 0.29  &  0.13  & 0.16     &   0.13      &         & 0.16$\pm0.05$ &  6950  &  -0.0 & 1,16 \\
10)IRAS 14325-6428          & 313.9  & +4.1   &  F5I   & 12.0   & 11.27  & 0.56  &  0.32  & 0.24     &  0.64       &0.89     & 0.89$\pm0.20$ &  6400  &  -0.03 & 1,11,12 \\
11)IRAS 15210-6554          & 317.7  & -7.7   &  K2I   & 11.85  & 11.72  & (0.03)*&  1.36 &          &  0.20      &         &     ...          &  4310  &  -0.61 & 1 \\
12)HD 137569                & 21.9   & +51.9  &  B9Iab:p&   7.91 & 7.89  & -0.05 & ...    & 0.0      &  0.05       & 0.01    & 0.01$\pm0.05$ &  10,500&  -0.53 & 1,17,18\\
13)BD+33 2642               & 52.7   & +50.8  &   O7p  & 10.73  & 10.78  & -0.12 &  -0.27 & 0.15     &      0.02  &  0.06   & 0.15$\pm0.05$ &  20,000&   -1.66 &1,19,20 \\
14)BD+32 2754               & 53.6   & +41.5  &   F8   & 9.55   & 9.46   & 0.57  & 0.56   & 0.01     &  0.02       & 0.02    & 0.01$\pm0.05$ &  5750  &   -0.09 & 1,14 \\
15)HD161796 &  77.1  &+30.9   &  F3Ib  & 7.21   & 7.08   & 0.47  & 0.26   & 0.21     &  0.03       & 0.04    & 0.21$\pm0.05$ &  6500  &   -0.03 & 1,9,12,21  \\
( IRAS 17436+5003)  &&&&&&&&&&&&& & \\
16)BD-12 4970               & 018.0  & +1.6   &  B0.5Ia&  8.78  & 8.30   & 1.02  & -0.21  & 1.23     &  2.71       & 1.58    & 1.23$\pm0.20$ &  27,000&   -2.40 & 1\\
17)IRAS 18075-0924          & 019.8  & +4.7   &  ----  &  13.9  & 12.47  & 1.4   &        &          &  1.41       &         & ...  &        & & 1\\
18)HD 178443                & 354.2  &  -21.5 &  F8    & 10.02  & 9.80   & 0.673  &  0.56 & 0.11     &  0.09       &         & 0.11$\pm0.05$ &  5180  &    -0.09 &1,22\\
19)PHL 1580                 & 031.3  &  -43.5 &  B0I   & 12.33  & 12.19  & (0.14)* &-0.22 &          &  0.04       &0.03     & 0.03$\pm0.05$ &  24,000&    -2.8 &1,23\\
20)LS III +52 5             & 095.1  &  +0.8  &  OB-   & (12.2)* & 11.74 & (0.46)* &-0.22 &          &  2.91      & 0.03    & 0.03$\pm0.05$ &  25,000&    -2.9 &1,24\\
\hline\noalign{\vskip 3pt} 
\end{tabular}
}
\begin{tabnote}
  Notes:- ()* indicates (V-R) for 11)IRAS 15210-6554, (V-G) for 19)PHL 1580, and B mag and (B-G) for 20)LS III +52 5. $E(B-V)_{\bf Sp}$ indicates $(B-V)-(B-V)_{0}$, where $(B-V)_{0}$ is estimated from the spectral type. $E(B-V)_{\rm SFD}$ is from the 2D dust extinction map of Schlegel et al. (1998). $E(B-V)_{\rm 3D}$ is taken from 3D dust maps of Green et al. (2019) if $|b|>10^\circ$ and Chen et al. (2019) if $|b|<10^\circ$. References: 1)SIMBAD; 2)\citet{fujii02}; 3)\citet{omont93}; 4)\citet{iyengar97}; 5)\citet{gielen11}; 6)\citet{rao12}; 7)\citet{oomen18}; 8)\citet{luck84}; 9)\citet{partha86}; 10)\citet{partha92}; 11)\citet{desmedt16}; 12)\citet{kamath22}; 13)\citet{gielen11}; 14)\citet{vickers15}; 15)\citet{reddy96}; 16)\citet{gonzalez92}; 17)\citet{martin04}; 18)\citet{martin06}; 19)\citet{napiwotzki94}; 20)\citet{vanwinckel14}; 21)\citet{luck90}; 22)\citet{mcwilliam95}; 23)\citet{mccausland92}; 24)\citet{hardorp64}
  \end{tabnote}
\end{table*}

\begin{table*}
\tbl{Gaia DR2  parallaxes and  derived luminosities of twenty high velocity post-AGB candidates
\label{tab:param}}{
\begin{tabular}{lrrrrrrrl}
\hline\noalign{\vskip3pt}
Star           &     parallax\footnotemark[$*$]   &  Distance &  Distance (BJ) & $\log(L/L_{\odot})$\footnotemark[$\ddagger$] &   $\log$($T_{\rm eff}$/K)  &  RV\footnotemark[$\dagger$] & RUWE & Subsection \\
 & (mas) & (kpc) & (kpc) & & & (km~s$^{-1}$) & & in Sect. 3 \\
\hline\noalign{\vskip3pt}
1)IRAS 02143+5852 &      1.364$\pm$0.289  & $>$0.510           &    &  $>$0.74       & 3.778 & -49.02$\pm$14.69   & 18.689 &3.4 \\
2)IRAS 05089+0459 &      0.754$\pm$0.345  & $>$0.684           &    &  $>$0.88       & 3.477 &  85.92$\pm$1.49    & 21.908 &3.4 \\
3)IRAS 05208-2035 &      0.687$\pm$0.030  &  1.420$\pm$0.064 & 1.403 $^{+0.053}_{-0.059}$ & 2.91$\pm$0.25  & 3.690 &  52.84$\pm$3.68        &  2.184 &3.4 \\
4)HD 46703        &      0.268$\pm$0.024  &  3.512$\pm$0.330 & 3.399 $^{+0.276}_{-0.278}$ & 3.92$\pm$0.26  & 3.778 &  -83.53$\pm$7.71       & 1.622  &3.2 \\
5)HD 56126        &      0.454$\pm$0.024  &  2.124$\pm$0.114 & 2.099 $^{+0.108}_{-0.110}$ & 3.93$\pm$0.25  & 3.813 & 93.71$\pm$3.54         & 0.922  &3.1  \\
6)IRAS 07140-2321 &      0.178$\pm$0.012  &  5.116$\pm$0.343 & 5.122 $^{+0.377}_{-0.404}$ & 3.74$\pm$0.26  & 3.845 &  62.38$\pm$4.13        & 1.029  &3.1  \\
7)IRAS 07227-1320 &      0.489$\pm$0.021  &  1.975$\pm$0.087 & 1.982 $^{+0.068}_{-0.073}$ & 2.58$\pm$0.25  & 3.544 &  70.68$\pm$0.39        & 1.176  &3.3  \\
8)IRAS 08187-1905 &      0.288$\pm$0.033  &  3.280$\pm$0.403 & 3.259 $^{+0.342}_{-0.391}$ & 3.60$\pm$0.12  & 3.789 & 65.44$\pm$1.87         & 1.695  &3.2 \\
9)HD 116745       &      0.177$\pm$0.020  &  5.154$\pm$0.597 & 4.893 $^{+0.379}_{-0.444}$ & 3.20$\pm$0.11  & 3.842 & 240.11$\pm$0.54        & 0.933  &3.1  \\
10)IRAS 14325-6428 &      0.192$\pm$0.037  &  4.795$\pm$1.033 & 4.883 $^{+0.622}_{-0.928}$ & 2.77$\pm$0.30  & 3.806 &-76.54$\pm$10.08        & 2.181  &3.2 \\
11)IRAS 15210-6554 &     -0.152$\pm$0.143&   $>$6.623           &  &   $>$3.29            & 3.634 &-83.90$\pm$0.87 & 2.279 & 3.4 \\
12)HD 137569       &      0.752$\pm$0.079  &  1.301$\pm$0.150 & 1.316 $^{+0.132}_{-0.197}$ & 3.19$\pm$0.11  & 4.021 & -45.0                  & 2.023 &3.2 \\
13)BD+33 2642      &      0.271$\pm$0.032  &  3.474$\pm$0.434 & 3.467 $^{+0.308}_{-0.466}$ & 3.54$\pm$0.12  & 4.301 &-94.7$\pm$2.5           & 1.295 &3.1 \\
14)BD+32 2754      &      3.239$\pm$0.014  &  0.307$\pm$0.001 & 0.307 $^{+0.001}_{-0.001}$ & 1.10$\pm$0.06  & 3.760 & -60.50$\pm$0.50        & 1.161 &3.3  \\
15)HD 161796 &      0.502$\pm$0.024  &  1.926$\pm$0.091 & 1.921 $^{+0.091}_{-0.095}$& 3.85$\pm$0.07  & 3.813 & -54.17$\pm$1.78               & 1.216 &3.1  \\
16)IRAS 18075-0924 &    -0.171$\pm$0.192    &  $>$4.348         &     &                 &       & -59.68$\pm$0.68   & 7.580 & 3.4 \\
17)BD-12 4970      &      0.467$\pm$0.020  &  2.065$\pm$0.090 & 1.984 $^{+0.072}_{-0.101}$& 5.50$\pm$0.25  & 4.431 &  124.95$\pm$9.43       & 0.956 &3.1   \\
18)HD 178443       &      1.034$\pm$0.016  &  0.951$\pm$0.014 & 0.939 $^{+0.014}_{-0.015}$ & 1.99$\pm$0.06  & 3.714 &  343.55$\pm$0.28       & 1.102 &3.3   \\
19)PHL 1580        &      3.156$\pm$0.018  &  0.315$\pm$0.002 & 0.314 $^{+0.001}_{-0.002}$ & 1.12$\pm$0.06  & 4.380 & -70.53$\pm$0.72        & 0.954 &3.3   \\
20)LS III +52 5    &      3.119$\pm$0.011  &  0.319$\pm$0.001 & 0.315 $^{+0.001}_{-0.001}$ & 1.22$\pm$0.06 & 4.398 &-232.83$\pm$0.67         & 0.817 &3.3   \\
\hline\noalign{\vskip 3pt} 
\end{tabular}
}
\begin{tabnote}
\footnotemark[$*$] From Gaia EDR3 (Lindegren et al. A\&A, 2021, in press. arxiv:2012.03380).\\
\footnotemark[$\dagger$] From Gaia DR2 except for IRAS 07140--2321, HD 137569, and BD+33~2642, for which the values are respectively taken from RAVE DR6 (Steinmetz et al. 2020, AJ, 160, 82), Dulflot et al. (1995, \aaps,114, 269), and Gontcharov, G.~A.(2006, Astronomy Letters, 32, 759).\\
\footnotemark[$\ddagger$] The luminosity uncertainty includes the uncertainty in distance and reddening (\S~2). In case the relative parallax measurement uncertainty is larger than 20$\%$, we provide $2\sigma$ lower limits.
\end{tabnote}
\end{table*}

\begin{table}
\tbl{Kinematics information of twenty high velocity post-AGB candidates
\label{tab:kinematics}}{
\begin{tabular}{lrrrrrr}\hline
 & $v_{T}$\footnotemark[$*$] & $v_{\phi}$  & $v_R$  & $v_z$ & $L_z$ & $E$ \\
 &  (km~s$^{-1}$) &  (km~s$^{-1}$) &  (km~s$^{-1}$) &  (km~s$^{-1}$) &  (kpc km~s$^{-1}$) & (km$^{2}$~s$^{-2}$) \\
 \hline
1)IRAS 02143+5852      &   $>$8.6 &     &   &   &    &      \\
2)IRAS 05089+0459      &   $>$18.5 &     &    &     &    &       \\
3)IRAS 05208-2035      &   16.2 &  216.3 &  13.0 &   -4.9 &   1983.8 &   -153156 \\
4)HD 46703             &   66.5 &  162.6 & -51.8 &  -17.7 &   1857.2 &   -149824 \\
5)HD 56126              &    5.1 &  207.8 &  52.1 &   18.9 &   2104.9 &   -148503 \\
6)IRAS 07140-2321      &   63.0 &  240.1 & -11.9 &   -3.8 &   2839.3 &   -134521 \\
7)IRAS 07227-1320      &   15.0 &  206.9 &  15.1 &    8.0 &   1992.9 &   -152988 \\
8)IRAS 08187-1905      &   36.9 &  207.3 &  -8.7 &   -1.0 &   2119.8 &   -149469 \\
9)HD 116745            &  185.4 &  -84.6 & -56.6 &  -74.8 &   -541.3 &   -184843 \\
10)IRAS 14325-6428      &  207.3 &  241.2 &  55.9 &   44.0 &   1457.7 &   -167529 \\
11)IRAS 15210-6554      &  $>$273.0 &     &    &     &       &       \\
12)HD 137569            &   90.9 &  226.4 & -50.2 &  -79.3 &   1688.5 &   -156543 \\
13)BD+33 2642           &  238.3 &   13.5 & 142.2 &   85.1 &     98.0 &   -169581 \\
14)BD+32 2754           &   66.9 &  159.4 &  20.9 &   12.8 &   1286.9 &   -170975 \\
15)HD 161796            &  109.6 &  193.8 & -69.7 &  -17.5 &   1551.1 &   -161792 \\
16)BD-12 4970           &   27.0 &  272.0 &-111.3 &    3.6 &   1706.0 &   -154487 \\
17)IRAS 18075-0924      & $>$137.4 &     &    &     &       &       \\
18)HD 178443            &  233.1 &  -21.9 &-300.4 & -131.0 &   -160.5 &   -135111 \\
19)PHL 1580             &   47.6 &  186.6 &  51.6 &   24.1 &   1495.6 &   -165370 \\
20)LS III +52 5         &   79.7 &    7.0 &  33.3 &  -42.0 &     57.6 &   -181508 \\\hline
\end{tabular}
}
\begin{tabnote}
\footnotemark[$*$] Tangential velocity computed from the proper motion and parallax. In case the relative parallax measurement uncertainty is larger than 20\%, we provide 2$\sigma$ lower limit.
\end{tabnote}
\end{table}

\end{document}